\documentstyle[prd,aps]{revtex}

\preprint{
\text{LAVAL-PHY-97-12},
\text{hep-ph}
}

\begin{document}
\author{F. Leblond and L. Marleau}
\address{D\'epartement de Physique, Universit\'e Laval\\
Qu\'ebec, Canada, G1K 7P4}
\title{Oblate Skyrmions}
\date{July 1997}
\maketitle

\begin{abstract}
The spherically symmetric hedgehog ansatz used in the description of the
skyrmion is believed to be inadequate for the rotational states such as the
nucleon ($I=J=\frac{1}{2}$) and the $\Delta $ ($I=J=\frac{3}{2}$) due to
centrifugal forces. We study here a simple alternative: an oblate spheroidal
solution which leads to lower masses for these baryons. As one might expect,
the shape of the solution is flatter as one increases $I=J$ whether the size
of the soliton is allowed to change or not.
\end{abstract}

\pacs{PACS numbers: 12.39.Dc, 11.30.Na. }

\section{Introduction}

When Skyrme first introduced its model a few decades ago \cite{Skyrme61} to
describe baryons as solitons in a non-linear field theory of mesons, the
solution proposed was in the spherically symmetric hedgehog ansatz. There
are reasons to believe that this solution is not adequate for the rotational
states such as the nucleon ($I=J=\frac{1}{2}$) and the $\Delta $ ($I=J=\frac{%
3}{2}$) due to centrifugal forces \cite{Blaizot88,Li87,Wambach87}.
Alternative treatments have been proposed in the past with relative success.
These approaches generally fall into three classes: (a) The original
spherical shape of the solution is modified. This is usually done by making
a global deformation along one or more axis \cite{Hadjuk84}. (b) The size of
the skyrmion is allowed to change. This analysis also led to the
identification of breathing modes \cite{Hadjuk84,Biedenharn85} with excited
states of the nucleon and $\Delta $-isobar. Combining deformations (a) and
(b), one gets the following scheme: the nucleon's and $\Delta $-isobar's
ground states have ${\bf K}={\bf I+J}=0$ and spherical symmetry. This led to
the conclusion that these states are stable against quadrupole deformation 
\cite{Wambach87,Hadjuk84}. The $K\neq 0$ states occur in nearly degenerate
doublets for all spin: one oblate and one prolate. (c) Finally, the shape of
the solution itself could be improved by including the (iso-) rotational
kinetic energy in the minimization of the static Hamiltonian. It turns out
that there is no finite static hedgehog solution unless one considers
massive pions \cite{Braaten85,Liu84}. This instability is understood to come
from the emission of pions from a rapidly rotating skyrmion. Some progress
has been made recently to elucidate the connection between skyrmions and
Feynman diagrams in an effective field theory but it is also interesting to
note that the solution were characterized by small quadrupole deformations
away from the spherical hedgehog ansatz \cite{Schroers94,Dorey94}.

In this work, we take a naive approach and propose a simple alternative.
Instead of the spherically symmetric hedgehog solution, we introduce an
oblate spheroidal solution. This leads to lower masses and quadrupole
deformations for these baryons. Moreover, the shape of the solution is
flatter as one increases $I=J$ whether one allows the size of the soliton to
change or not.

In the next section, we introduce the oblate spheroidal coordinates. We then
proceed as follows: We determine the profile $f\left( \eta \right) $ by
solving the differential equation obtained from minimization of the static
energy. This procedure is similar to that of ref. \cite{Adkins83} and one
indeed recovers the profile of the hedgehog ansatz in the limit of parameter 
$d\rightarrow 0$. In Section III, we compute the masses of the nucleon ($I=J=%
\frac{1}{2}$) and of the $\Delta $-isobar ($I=J=\frac{3}{2}$). These masses
get contributions both from the static and rotational energy and will in
general depend on the choice of $d$. The value of $d$ for each baryon is
fixed by minimizing its mass with respect to $d$. In the last section we
discuss our results with respect to scale transformations and check that the
above scheme persists.

\section{The static oblate soliton}

The oblate spheroidal coordinates $\left( \eta ,\theta ,\phi \right) $ are
related to Cartesian coordinates through the expressions 
\begin{eqnarray}
x &=&d\cosh \eta \sin \theta \cos \phi  \nonumber \\
y &=&d\cosh \eta \sin \theta \sin \phi \\
z &=&d\sinh \eta \cos \theta .  \nonumber
\end{eqnarray}
A surface of constant $\eta $ corresponds to a sphere of radius $d$
flattened in the $z$-direction by a factor of $\tanh \eta $. For $\eta $
small, the shape of the surface is more like that of a pancake of radius $d$
whereas for large $\eta $, one recovers a spherical shell of radius $r=\frac{%
de^{\eta }}{2}$. Let us note that in the limit $d\rightarrow 0$, $\eta
\rightarrow \infty $ with $r$ remaining finite, the coordinate system
becomes 
\[
(x,y,z)=r(\sin \theta \cos \phi ,\sin \theta \sin \phi ,\cos \theta ) 
\]
which means that it coincides with the spherical coordinate system. The
choice of the parameter $d$ determines at what scale the ``oblateness''
becomes important. The element of volume is given by 
\begin{equation}
dV=-d^{3}\left( \cosh \eta \right) \left( \cosh ^{2}\eta -1+\cos ^{2}\theta
\right) \cdot d\eta d(\cos \theta )d\phi
\end{equation}

We would like to replace the hedgehog solution for the Skyrme model by an
oblate solution. Writing the Lagrangian for the Skyrme model \cite{Adkins83} 
\begin{equation}
{\cal L}=-\frac{F_{\pi }^{2}}{16}Tr\left( L_{\mu }L^{\mu }\right) +\frac{1}{%
32e^{2}}Tr\left( \left[ L_{\mu },L_{\nu }\right] ^{2}\right)
\end{equation}
where $L_{\mu }=U^{\dagger }\partial _{\mu }U$ with $U\in SU(2)$. We get the
usual expression for static energy density 
\begin{eqnarray}
{\cal E} &=&{\cal E}_{2}+{\cal E}_{4}  \nonumber \\
&=&-\frac{F_{\pi }^{2}}{16}Tr\left( L_{i}L_{i}\right) -\frac{1}{32e^{2}}%
Tr\left( \left[ L_{i},L_{j}\right] ^{2}\right) .
\end{eqnarray}

Let us now define a static oblate solution by 
\begin{equation}
U=e^{i\left( {\bf \tau \cdot \hat{\eta}}\right) f\left( \eta \right) }\ 
\label{oblate}
\end{equation}
where$\ {\bf \hat{\eta}}$ is the unit vector$\ {\bf \hat{\eta}}=\frac{{\bf %
\nabla }\eta }{\left| {\bf \nabla }\eta \right| }$. The boundary conditions
for the winding number $N=1$ solution are $f\left( 0\right) =\pi $ and $%
f\left( \infty \right) =0$. Note that this is not a priori a solution of the
field equations derived from the Skyrme Lagrangian.

Using the oblate ansatz $U=\exp \left[ i\left( {\bf \tau \cdot \hat{\eta}}%
\right) f\left( \eta \right) \right] $ we get, after a straightforward but
tedious calculation and an integration over the angular variables $\theta $
and $\phi ,$ the static energy contributions $E_{2}$ and $E_{4}$ such that 
\begin{equation}
E_{2}=\int dV{\cal E}_{2}=\frac{4\pi \epsilon }{\lambda }\cdot \frac{%
\widetilde{d}}{2}\cdot \int_{0}^{\infty }d\eta \left( \alpha _{21}f^{\prime
2}+\alpha _{22}\sin ^{2}f\right)  \label{E2}
\end{equation}
with 
\begin{eqnarray*}
\alpha _{21}\left( \eta \right) &=&2\cosh \eta \\
\alpha _{22}\left( \eta \right) &=&2\left( -2\cosh \eta +\left( 2\cosh
^{2}\eta -1\right) L\left( \eta \right) \right)
\end{eqnarray*}
where $L\left( \eta \right) \equiv \ln \left( \frac{\cosh \eta +1}{\cosh
\eta -1}\right) $ and 
\begin{equation}
E_{4}=\int dV{\cal E}_{4}=\frac{4\pi \epsilon }{\lambda }\cdot \frac{1}{4%
\widetilde{d}}\cdot \int_{0}^{\infty }d\eta \left( \alpha _{41}f^{\prime
2}\sin ^{2}f+\alpha _{42}\sin ^{4}f\right)  \label{E4}
\end{equation}
with 
\begin{eqnarray*}
\alpha _{41}\left( \eta \right) &=&2L\left( \eta \right) \\
\alpha _{42}\left( \eta \right) &=&\frac{1}{2}\left( \frac{1}{\cosh ^{2}\eta 
}\left( 2\cosh \eta +L\left( \eta \right) \right) +\frac{2\cosh \eta }{%
\left( \cosh ^{2}\eta -1\right) }\right)
\end{eqnarray*}
Here, we have expressed the parameters of the model in terms of the
constants: 
\begin{equation}
\epsilon =\frac{1}{\sqrt{2}e}\qquad \lambda =\frac{2}{F_{\pi }}\qquad 
\widetilde{d}=\frac{eF_{\pi }}{2\sqrt{2}}d.
\end{equation}

It might be required to add a pion mass term to the Skyrme Lagrangian. This
term takes the usual form 
\begin{equation}
{\cal L}_{m}=\frac{m_{\pi }^{2}F_{\pi }^{2}}{8}\left( TrU-2\right)
\end{equation}
leading to the expression 
\begin{equation}
E_{m_{\pi }}=\frac{4\pi \epsilon }{\lambda }\cdot 32\epsilon ^{2}\epsilon
_{\pi }^{2}\widetilde{d}^{3}\cdot \int_{0}^{\infty }d\eta \alpha _{m}\left(
1-\cos f\right)  \label{Empi}
\end{equation}
where $\epsilon _{\pi }=\frac{m_{\pi }}{F_{\pi }}$ and 
\[
\alpha _{m}\left( \eta \right) =\cosh \eta \left( \sinh ^{2}\eta +\frac{1}{3}%
\right) . 
\]
Minimizing the static energy with respect to $f\left( \eta \right) $, we
need to solve (numerically of course) the non-linear ordinary second-order
differential equation which reads: 
\begin{eqnarray}
0 &=&32\epsilon ^{2}\epsilon _{\pi }^{2}\widetilde{d}^{3}\cdot \alpha
_{m}\sin f+\frac{\widetilde{d}}{2}\cdot \left( -2f^{\prime \prime }\alpha
_{21}-2f^{\prime }\alpha _{21}^{\prime }+2\alpha _{22}\sin f\cos f\right) 
\nonumber \\
&&+\frac{1}{4\widetilde{d}}\cdot \left( \left( 4\sin ^{3}f\cos f\right)
\alpha _{42}+\left( -2f^{\prime \prime }\sin ^{2}f-2f^{\prime 2}\sin f\cos
f\right) \alpha _{41}-\left( 2f^{\prime }\sin ^{2}f\right) \alpha
_{41}^{\prime }\right) .  \label{EqChirale}
\end{eqnarray}

For calculational purposes, we need to set the value of the parameters of
the Skyrme Model. $F_{\pi }$ and $e$ are first chosen to coincide with those
of ref. \cite{Adkins83}:

\begin{eqnarray}
F_{\pi } &=&129\text{ MeV}\hspace{0.25in}e=5.45\hspace{0.25in}m_{\pi }=0
\label{ANW} \\
F_{\pi } &=&108\text{ MeV}\hspace{0.25in}e=4.84\hspace{0.25in}m_{\pi }=138%
\text{ MeV}  \label{AN}
\end{eqnarray}
obtained by fitting for the masses of the nucleon and the $\Delta $ in the
hedgehog ansatz.

The solution near $\eta \rightarrow 0$ has the form 
\[
f\left( \eta \right) \sim \pi -a_{1}\eta 
\]
whereas in the limit $\eta \rightarrow \infty $, one recovers the spherical
symmetry with, 
\begin{equation}
f\left( \eta \right) \sim k\left[ \frac{2m_{\pi }}{de^{\eta }}+\frac{4}{%
(de^{\eta })^{2}}\right] \exp \left( -\frac{m_{\pi }de^{\eta }}{2}\right)
\label{SolutionInf}
\end{equation}
where $a_{1}$ and $k$ are constants which depend on $\widetilde{d}$ and $%
m_{\pi }$. The solutions of differential equation (\ref{EqChirale}) are
presented for several values of $\widetilde{d}$ and $m_{\pi }=0$ in Fig. \ref
{Profile}. For small $\widetilde{d}$ (here $\widetilde{d}\leq 0.0001$), we
get exactly the solution of the spherical hedgehog skyrmion. As one
increases $\widetilde{d},$ we observe a displacement of the function $%
f\left( \eta \right) $ and the continuous deformation of the soliton from a
spherical to an oblate shape. The static energy, $E_{s}=E_{m_{\pi
}}+E_{2}+E_{4}$, has a minimum for $\widetilde{d}=0$ which is expected since
it corresponds to the spherical solution (see Fig. \ref{FigureEnergies}).

The masses of the nucleon and of the $\Delta $-isobar get contributions both
from the static and rotational energy and will generally depend on the
choice of $\widetilde{d}$. We fix the value of $\widetilde{d}$ for each
baryon by minimizing its mass with respect to $\widetilde{d}$.

\section{Collective variables}

Using the oblate solution, we can then compute the masses of the nucleons ($%
I=J=\frac{1}{2}$) and of the $\Delta $-isobar ($I=J=\frac{3}{2}$). However,
several remarks are in order before we go on. When one departs from the
spherical symmetry of the hedgehog ansatz, it is customary to introduce
extra collective variables for isorotation in addition to those
characterizing spatial rotation since these are no longer equivalent, in
general. The spin and isospin contributions to the rotational energy are
however equal in our case since we use solution (\ref{oblate}) and we are

only interested in ground states with ${\bf K}={\bf J}+{\bf I}=0$ (see
Appendix). As a result, we need only consider one set of collective
variables.

Let us work in the body-fixed system and assume that the time dependence can
be introduced using the usual substitution 
\begin{equation}
U\rightarrow A(t)UA^{\dagger }(t)  \label{collective}
\end{equation}
where $A(t)$ is a time-dependent $SU(2)$ matrix. This transformation leaves
the static energy (or mass of the soliton) invariant. We can then go on and
treat $A(t)$ approximately as quantum mechanical variables. The calculation
procedure is fairly standard (see ref. \cite{Adkins83} for example).

Using (\ref{collective}), the Lagrangian gets new terms due to the time
dependence of $A$:

\begin{equation}
L_{2}^{t}=\int dV{\cal L}_{2}^{t}=-\frac{F_{\pi }^{2}}{16}\int dV\ \text{Tr}%
\left( \widetilde{L}_{0}\widetilde{L}^{0}\right)
\end{equation}
and 
\begin{equation}
L_{4}^{t}=\int dV{\cal L}_{4}^{t}=-\frac{1}{32e^{2}}\int dV\ \text{Tr}\left(
\left[ \widetilde{L}_{0},L_{i}\right] ^{2}\right)
\end{equation}
where 
\begin{equation}
\widetilde{L}_{0}=AU^{\dagger }A^{\dagger }\partial _{0}\left( AUA^{\dagger
}\right)
\end{equation}

Following straightforward but lengthy calculations, we get after angular
integrations 
\begin{equation}
L_{2}^{t}=\frac{1}{2}a_{2}^{ij}Tr\left[ \tau _{i}A^{\dagger }\dot{A}\right]
Tr\left[ \tau _{j}A^{\dagger }\dot{A}\right]
\end{equation}
and

\begin{equation}
L_{4}^{t}=-\frac{1}{2}a_{4}^{ij}Tr\left[ \tau _{i}A^{\dagger }\dot{A}\right]
Tr\left[ \tau _{j}A^{\dagger }\dot{A}\right]
\end{equation}
where 
\begin{eqnarray}
a_{2}^{ij} &=&\frac{\lambda }{4\pi \epsilon }\cdot 128\pi ^{2}\epsilon ^{4}%
\widetilde{d}^{2}\cdot \frac{\widetilde{d}}{2}\int_{0}^{\infty }d\eta \cosh
\eta \sin ^{2}f\ A_{ij}  \nonumber \\
a_{4}^{ij} &=&\frac{\lambda }{4\pi \epsilon }\cdot 128\pi ^{2}\epsilon ^{4}%
\widetilde{d}^{2}\cdot \frac{1}{4\widetilde{d}}\int_{0}^{\infty }d\eta \cosh
\eta \sin ^{2}f\ \left[ C_{ij}\sin ^{2}f+B_{ij}f^{\prime 2}\right]
\label{aij}
\end{eqnarray}
with 
\begin{eqnarray}
A_{11} &=&A_{22}=\left( \cosh ^{2}\eta -\frac{1}{2}\right) \left( -\frac{2}{3%
}\left( 3\cosh ^{2}\eta -4\right) +\cosh \eta \sinh ^{2}\eta \ L\left( \eta
\right) \right)  \nonumber \\
A_{33} &=&4\cosh ^{4}\eta -\frac{10}{3}\cosh ^{2}\eta -\left( 2\cosh
^{5}\eta -3\cosh ^{3}\eta +\cosh \eta \right) L\left( \eta \right)  \nonumber
\\
B_{11} &=&B_{22}=2-\cosh ^{2}\eta -\frac{1}{2}\cosh \eta \left( 1-\cosh
^{2}\eta \right) L\left( \eta \right)  \nonumber \\
B_{33} &=&2\cosh ^{2}\eta +\cosh \eta \left( 1-\cosh ^{2}\eta \right)
L\left( \eta \right) \\
C_{11} &=&C_{22}=\left( \frac{-1}{4\cosh \eta }\right) \left( 10\cosh \eta
-8\cosh ^{3}\eta +\left( 4\cosh ^{4}\eta -9\cosh ^{2}\eta +3\right) L\left(
\eta \right) \right)  \nonumber \\
C_{33} &=&\frac{1}{2\cosh \eta }\left( \left( 4\cosh ^{4}\eta -\cosh
^{2}\eta -1\right) L\left( \eta \right) -8\cosh ^{3}\eta +2\cosh \eta
\right) .  \nonumber
\end{eqnarray}
Non-diagonal terms for $A_{ij}$, $B_{ij}$ and $C_{ij}$ give zero
contribution upon $\phi $ integration due to the axial symmetry of the
solution.

Let us now consider the quantity 
\[
Tr\left[ \tau _{i}A^{\dagger }\dot{A}\right] Tr\left[ \tau _{j}A^{\dagger }%
\dot{A}\right] a_{n}^{ij} 
\]
for an axially symmetric system where $a_{n}^{11}=a_{n}^{22}\neq a_{n}^{33}$
and $a_{n}^{ij}=0$ for $i\neq j$. We can rewrite 
\begin{equation}
Tr\left[ \tau _{i}A^{\dagger }\dot{A}\right] Tr\left[ \tau _{j}A^{\dagger }%
\dot{A}\right] a_{n}^{ij}=Tr\left( \dot{A}\dot{A}^{\dagger }\right)
a_{n}^{11}+\left( Tr\left[ \tau _{3}A^{\dagger }\dot{A}\right] \right)
^{2}\left( a_{n}^{33}-a_{n}^{11}\right) .
\end{equation}
In terms of the Euler angles $\Theta ,\Phi $ and $\Psi $, the traces
correspond to the expressions 
\begin{equation}
Tr\left( \dot{A}\dot{A}^{\dagger }\right) =\frac{1}{2}\left( \dot{\Theta}%
^{2}+\dot{\Phi}^{2}+\dot{\Psi}^{2}+2\cos \Theta \dot{\Phi}\dot{\Psi}\right)
\end{equation}
\begin{equation}
Tr\left[ \tau _{3}A^{\dagger }\dot{A}\right] =i\left( \dot{\Psi}+\cos \Theta 
\dot{\Phi}\right)
\end{equation}
which finally leads to 
\begin{eqnarray}
L_{2}^{t}+L_{4}^{t} &=&\frac{b}{2}\left( \dot{\Theta}^{2}+\sin ^{2}\Theta 
\dot{\Phi}^{2}\right) +\frac{c}{2}\left( \dot{\Psi}+\cos \Theta \dot{\Phi}%
\right) ^{2}  \nonumber \\
&=&\frac{b}{2}\Omega _{1}^{2}+\frac{c}{2}\Omega _{2}^{2}  \label{Lrot}
\end{eqnarray}
where 
\begin{eqnarray}
b &=&\left( a_{2}^{11}+a_{4}^{11}\right) \\
c &=&\left( a_{2}^{33}+a_{4}^{33}\right) .
\end{eqnarray}
Here $b$ and $c$ play the role of principal moment of inertia.

Quantization of (\ref{Lrot}) is straightforward. It indeed represents a
symmetrical top with the rotational kinetic energy in space and isospace 
\begin{equation}
E_{rot}^{J,J_{3}}=\frac{1}{2b}\left( \frac{\left| {\bf J}\right| ^{2}+\left| 
{\bf I}\right| ^{2}}{2}\right) +\frac{1}{2}\left( \frac{1}{c}-\frac{1}{b}%
\right) J_{3}^{2}.  \label{Bodyfixed}
\end{equation}
where $\left| {\bf J}\right| ^{2}$ and $\left| {\bf I}\right| ^{2}$ are the
spin and the isospin respectively and, $J_{3},$ the $z$-component of the
spin. We have already used the relation $J_{3}=-I_{3}$ here which follows
from axial symmetry. Added to the static energy $E_{s},$ it leads to the
total energy 
\begin{equation}
M^{J,J_{3}}=E_{s}+E_{rot}^{J,J_{3}}.  \label{Mjj3}
\end{equation}

Up to now, we have analyzed the rotational and isorotational kinetic energy
from the point of view of the body-fixed frame. Observables states, however,
must be eigenstates of $\left| {\bf J}\right| ^{2},$ $J_{3},$ $\left| {\bf I}%
\right| ^{2}$, $I_{3}$ with eigenvalues $J(J+1),$ $m_{J}$, $I(I+1),$ $m_{I}$
where the operators now refer to the laboratory system (as opposed to
body-fixed operators in (\ref{Bodyfixed}) and above). These eigenstates can
be represented by direct products of rotation matrices 
\begin{equation}
\left\langle \alpha ,\beta ,\gamma |J,m_{J},m\right\rangle \left\langle \rho
,\sigma ,\tau |I,m_{I},-m\right\rangle =D_{m_{J}m}^{J}\left( \alpha ,\beta
,\gamma \right) D_{m_{I}-m}^{I}\left( \rho ,\sigma ,\tau \right)
\label{WignerD}
\end{equation}
where $\left( \alpha ,\beta ,\gamma \right) $ and $\left( \rho ,\sigma ,\tau
\right) $ are, respectively, the Euler angles for the rotation and
isorotation from the body-fixed frame to the laboratory system. Since we
have axial symmetry in the body-fixed system where $J_{3}=-I_{3},$ the
quantum number denoted by $m$ must be opposite in sign in space and isospace
rotation matrices. It is convenient to label the basis by the sum of the
body-fixed spin and isospin, ${\bf K.}$ The eigenstates of ${\bf K}^{2}$ are
linear combinations of the basis states (\ref{WignerD}) with Clebsch-Gordan
coefficients $\left\langle J,m;I,-m|K,0\right\rangle $. The explicit
calculation of the energy of rotation requires in general the
diagonalization $E_{rot}^{J,J_{3}}$. (see ref. \cite{Hadjuk84} for more
details). Since we are only interested in the ground states here, i.e. the
nucleon and $\Delta $-isobar, we set ${\bf K}=0$ which simplifies much of
the above procedure.

We proceded with the case $m_{\pi }=0$. Numerically, the minimization of the
static energy for the spherical symmetric ansatz gives $E_{s}=\frac{4\pi
\epsilon }{\lambda }\cdot \left[ 8.20675\right] $, $M_{N}=\frac{4\pi
\epsilon }{\lambda }\cdot \left[ 8.906\right] $ and $M_{\Delta }=\frac{4\pi
\epsilon }{\lambda }\cdot \left[ 11.703\right] $. For the oblate spheroidal
ansatz, the solution for $f\left( \eta \right) $ is found from (\ref
{EqChirale}) and the parameter $\widetilde{d}$ is chosen in order to
minimize the mass of the corresponding baryon. In general, as $\widetilde{d}$
increases, the static energy $E_{s}$ deviates from its lowest energy value
given by the spherical hedgehog configuration. On the other hand, oblate
configurations have larger moment of inertia which tends to decrease the
rotational kinetic energy (see Fig. \ref{FigureEnergies}). The existence of
a non-trivial oblate spheroidal ground state for the nucleon and the $\Delta 
$-isobar, as it turns out, depends mostly on the relative importance of
static and rotational energy.

Our results are summarized in Table \ref{resultats}. We find that the ground
state for the nucleon is almost spherical but nonetheless oblate with $%
\widetilde{d}=0.0013$ thus exhibiting a small quadrupole deformation and a
slightly lower mass with respect to a spherical configuration. For the $%
\Delta $-isobar, the oblateness or quadrupole deformation is even more
important and accounts for a $4\%$ decrease in mass. We obtain a minimum for
the $\Delta $ mass for a value of $\widetilde{d}=0.32$ with $M_{\Delta }=%
\frac{4\pi \epsilon }{\lambda }\cdot \left[ 11.293\right] $.

Since the minimum of the ground state is affected by the oblate shape of the
solution, the parameters $F_{\pi }$ and $e$ as given in ref. \cite{Adkins83}
no longer reproduce the quantities they were designed to fit. However, the
existence and the form of an oblate ground state for baryons depends on the
precise value of $F_{\pi },$ $e$ and $\widetilde{d}$ through eq. (\ref
{EqChirale}). Therefore in order to fit for $M_{N}=939$ MeV and $M_{\Delta
}=1232$ MeV, we must readjust $F_{\pi }$ and $e$ which determine the value
of $\widetilde{d}$ for the nucleon and $\Delta $-isobar respectively. After
several iterations, we find $F_{\pi }=118.4$ MeV and $e=5.10$ with $%
\widetilde{d}=0.0014$ ($\widetilde{d}=0.40$) for the nucleon ($\Delta $%
-isobar).

The numerical calculations for $m_{\pi }\neq 0$ lead to similar conclusions.
Starting from input values for $F_{\pi }$ and $e$ in (\ref{AN}), we get a
deformation parameter of $\widetilde{d}=0.0009$ for the nucleon and $%
\widetilde{d}=0.18$ for the $\Delta $-isobar leading to small decreases in
their respective masses. The deformation parameters here are significantly
smaller than what is observed in the $m_{\pi }=0$ case, which is partly
explained by the sensitivity of $\widetilde{d}$ with respect to $F_{\pi }$
and $e$. But since the chiral symmetry breaking term contributes here (i.e. $%
E_{m_{\pi }}$), this is also connected to the relative importance of the
rotational energy contribution to the baryon mass and perhaps more
importantly to how each contribution depends on $\widetilde{d}$.

\section{Discussion}

Quadrupole deformations were found previously \cite{Schroers94,Dorey94} in
the context of rotationally improved skyrmions. Contrary to our variational
approach, these solutions involve the minimization of a Hamiltonian which
includes the (iso-) rotational kinetic energy, i.e. eq. (\ref{EqChirale})
with contributions from the (iso) rotational kinetic energy. Nonetheless, we
found that the oblate spheroidal ansatz gives lower energy than the
spherical one for baryon ground states.

Of course, ansatz (\ref{oblate}) is not necessarily the lowest energy
solution, the latter being obtained in principle by solving the
integro-differential equation of ref.\cite{Dorey94}. Unfortunately, only
large-distance asymptotics of this solution can be written in a closed form.
Moreover, the most relevant physical quantity here, the mass of the baryons,
gets negligible contributions from that region and so it is not very
sensitive to the exact form of the solution at large distances. Yet, it may
be interesting to consider how our results compare to large-distance
asymptotics of the solution given in ref. \cite{Dorey94} in which the pion
field reads 
\begin{eqnarray}
{\bf \pi }({\bf r,J})\stackrel{r\rightarrow \infty }{\longrightarrow } &&%
\frac{B}{\widetilde{{\bf J}}^{2}}\left\{ \left[ \frac{m_{\pi }}{r}+\frac{1}{%
r^{2}}\right] \exp \left( -m_{\pi }r\right) \left( \widetilde{{\bf J}}\cdot 
{\bf r}\right) \widetilde{{\bf J}}\right.   \nonumber \\
&&+\left. \left[ \frac{\sqrt{m_{\pi }^{2}-\widetilde{{\bf J}}^{2}}}{r}+\frac{%
1}{r^{2}}\right] \exp \left( -\sqrt{m_{\pi }^{2}-\widetilde{{\bf J}}^{2}}%
r\right) \left( \widetilde{{\bf J}}\times {\bf r\times }\widetilde{{\bf J}}%
\right) \right\} .  \label{risky}
\end{eqnarray}
Here $\widetilde{J}^{k}=J^{k}\left( I_{mk}\right) ^{-1}$ where $J^{k}$ and $%
I_{mk}$ are the Skyrmion classical angular momentum and moment of inertia
tensor respectively. As one might expect, we recover the hedgehog solution
form in the limit $\widetilde{{\bf J}}^{2}\rightarrow 0$ and $r\rightarrow
\infty .$ For $\widetilde{{\bf J}}^{2}\neq 0$, the second term in (\ref
{risky}) dominates which can be interpreted as a swelling of the Skyrmion
with the pion field pointing in a direction perpendicular to $\widetilde{%
{\bf J}}$ due to centrifugal forces. Unfortunately, the exact magnitude and
direction of $\widetilde{{\bf J}}$ can only be obtained by solving the full
integro-differential equation. On the other hand, for the oblate ansatz the
pion field takes the form ${\bf \pi }({\bf \eta })={\bf \hat{\eta}}f\left(
\eta \right) $ and coincide with the hedgehog solution in the limit of large
distances. The magnitude of $k\,$in (\ref{SolutionInf}) is found by solving (%
\ref{EqChirale}) and optimizing for the deformation parameter $\widetilde{d}$%
. Our numerical calculations show that $k$ increases slowly with $\widetilde{%
d}$, which suggests that the configuration of the Skyrmion at large
distances is hedgehog-like and swelling for increasing isospin. This is in
partial agreement with the qualitative features of (\ref{risky}). Be that as
it may, we recall that the purpose was mainly to look at possible
deformations at middle-range distances since this is where energy and
baryonic densities are the largest. Ansatz (\ref{oblate}) turned out to be a
rather simple, intuitive and efficient trial solution.

It may also be interesting to consider deformations of the oblate skyrmions
under scaling of the unitary transformations $U\left( {\bf r}\right) $ such
that 
\begin{equation}
U\left( {\bf r}\right) =U_{0}\left( \rho {\bf r}\right)
\label{ScalingTransf}
\end{equation}
to minimize the total energy of the nucleon and $\Delta $-isobar. This
corresponds to skyrmions which are allowed to change in size. Recall that
the previous calculations proposed a change in shape (oblate spheroidal vs
spherical). The treatment is straightforward and indeed very similar to that
of ref. \cite{Hadjuk84}. In our calculations the scale transformation is
equivalent to the substitution 
\begin{equation}
\widetilde{d}\rightarrow \frac{\widetilde{d}}{\rho }.  \label{Scalingd}
\end{equation}

Rewriting the expression for the masses in the body-fixed frame as: 
\begin{equation}
M^{J,J_{3}}=E_{m_{\pi }}+E_{2}+E_{4}+\frac{\left( J\left( J+1\right)
-J_{3}^{2}\right) }{2\left( a_{2}^{11}+a_{4}^{11}\right) }+\frac{J_{3}^{2}}{%
2\left( a_{2}^{33}+a_{4}^{33}\right) }
\end{equation}
we see that under the scaling transformation (\ref{ScalingTransf}) 
\begin{equation}
M^{J,J_{3}}(\rho )=\frac{E_{m_{\pi }}}{\rho ^{3}}+\frac{E_{2}}{\rho }+\rho
E_{4}+\frac{\rho ^{3}\left( J\left( J+1\right) -J_{3}^{2}\right) }{2\left(
a_{2}^{11}+\rho ^{2}a_{4}^{11}\right) }+\frac{\rho ^{3}J_{3}^{2}}{2\left(
a_{2}^{33}+\rho ^{2}a_{4}^{33}\right) }.  \label{Mjj3rho}
\end{equation}
The total energies $M_{N}(\rho )$ and $M_{\Delta }(\rho )$, computed in the
laboratory system, can be minimized with respect to the $\rho $ parameter,
i.e. to the energically favored size of the oblate skyrmion. The results are
shown in Table \ref{resultats} for both the oblate and spherical cases. The
baryon ground states are now swelled oblate solutions. Again, one should in
principle readjust the $F_{\pi }$ and $e$ parameters to fit the masses of
the nucleon and $\Delta $-isobar. It would also be interesting to readdress
the problem of breathing modes with these oblate skyrmions. This is a
problem for further research.

\acknowledgements

We are indebted to Dr. N.N. Scoccola and M.Paranjape for useful comments and
discussions.This work was supported in part by the Natural Sciences and
Engineering Research Council of Canada and by the Fonds pour la Formation de
Chercheurs et l'Aide \`{a} la Recherche du Qu\'{e}bec.

\appendix

\section*{}%

The rotational energy for an axially symmetric system is given by 
\begin{eqnarray*}
L^{t} &=&\frac{A}{2}\left( \omega _{1}^{2}+\omega _{2}^{2}\right) +\frac{B}{2%
}\left( \Omega _{1}^{2}+\Omega _{2}^{2}\right) +\frac{C}{2}\left( \omega
_{3}-\Omega _{3}\right) ^{2} \\
&&-D\left( \omega _{1}\Omega _{1}+\omega _{2}\Omega _{2}\right)
\end{eqnarray*}
where ${\bf \omega }$ and ${\bf \Omega }$ are the angular velocities in
coordinate and isospin space. $A,B,C$ and $D$ are positive quantities
corresponding to moments of inertia. The previous expression can be written
in terms of body-fixed angular momentum ${\bf J}$ and isospin ${\bf I}$: 
\[
L^{t}=\frac{1}{AB-D^{2}}\left[ A\left( I_{1}^{2}+I_{2}^{2}\right) +B\left(
J_{1}^{2}+J_{2}^{2}\right) +2D\left( I_{1}J_{1}+I_{2}J_{2}\right) \right] +%
\frac{J_{3}^{2}}{C} 
\]
where $I_{i}$ and $J_{i}$ are the body-fixed components of ${\bf J}$ and $%
{\bf I}$. Here we have already used the fact that $J_{3}=-I_{3}$ because of
axial symmetry. Clearly, the nucleon and $\Delta $-isobar ground states are
obtained when ${\bf K}={\bf J}+{\bf I}=0$ since the term $2D\left(
I_{1}J_{1}+I_{2}J_{2}\right) $ is then negative. The general expressions for 
$A,B,C$ and $D$ are rather lengthy and therefore not given here (see ref. 
\cite{Sanyuk93}). However, they become much simpler for a solution of the
form (\ref{oblate}) in which case $A=B=D$ such that the rotational energy of
the nucleon and $\Delta -$isobar ground states are given by 
\[
L^{t}=\frac{1}{2A}\left( \left| {\bf J}\right| ^{2}-J_{3}^{2}\right) +\frac{1%
}{2C}J_{3}^{2}+\frac{1}{2A}\left( \left| {\bf I}\right|
^{2}-I_{3}^{2}\right) +\frac{1}{2C}I_{3}^{2}. 
\]
The spin and isospin contributions to the rotational energy are equal in
this case.

\newpage 
\begin{figure}[tbp]
\caption{Solutions of differential equation (\ref{EqChirale}) for several
values of $\widetilde{d}$ ($F_{\pi }=129$ MeV, $e=5.45$ and $m_{\pi }=0$).
For $\widetilde{d}=0.0001$, we get exactly the solution of the spherical
hedgehog skyrmion.}
\label{Profile}
\end{figure}

\begin{figure}[tbp]
\caption{Static and rotational energies for the nucleon as a function of $%
\widetilde{d}$ in units of $\frac{4\pi \epsilon }{\lambda }$. }
\label{FigureEnergies}
\end{figure}

\begin{table}[tbp]
\caption{Ground states for the nucleon and $\Delta $-isobar. The results are
shown for both the minimum oblate spheroidal configuration and the
spherically symmetric ansatz for comparison. The values of $M^{J,J_{3}}$ are
defined according to eq. (\ref{Mjj3}) whereas $M^{J,J_{3}}(\rho _{\min })$
is minimized with respect to the scaling parameter $\rho $ (see eq. (\ref
{Mjj3rho})). All masses are expressed in units of $\frac{4\pi \epsilon }{%
\lambda }$ with parameters $F_{\pi }=129$ MeV, $e=5.45$ and $m_{\pi }=0$. }
\label{resultats}
\begin{tabular}{ccccccccccccc}
& \qquad & \multicolumn{5}{c}{Oblate ($\widetilde{d}>0$)} & \qquad & 
\multicolumn{5}{c}{Spherical ($\widetilde{d}=0$)} \\ \cline{3-7}\cline{9-13}
&  & $\widetilde{d}$ & $M^{J,J_{3}}$ & \qquad & $\rho _{\min }$ & $%
M^{J,J_{3}}(\rho _{\min })$ &  &  & $M^{J,J_{3}}$ & \qquad & $\rho _{\min }$
& $M^{J,J_{3}}(\rho _{\min })$ \\ \cline{1-1}\cline{3-7}\cline{9-13}
Nucleon &  & \multicolumn{1}{l}{$0.0013$\quad} & \multicolumn{1}{r}{$8.904$}
&  & \multicolumn{1}{l}{$0.868$} & \multicolumn{1}{r}{$8.797$} &  &  & 
\multicolumn{1}{r}{$8.906$} &  & \multicolumn{1}{l}{$0.867$} & 
\multicolumn{1}{r}{$8.799$} \\ 
$\Delta $ &  & \multicolumn{1}{l}{$0.32$} & \multicolumn{1}{r}{$11.312$} & 
& \multicolumn{1}{l}{$0.670$} & \multicolumn{1}{r}{$10.064$} &  &  & 
\multicolumn{1}{r}{$11.703$} &  & \multicolumn{1}{l}{$0.668$} & 
\multicolumn{1}{r}{$10.238$}
\end{tabular}
\end{table}


\begin{references}
\bibitem{Skyrme61}  T.H.R. Skyrme, Proc. R. Soc. London {\bf A260}, 127
(1961).

\bibitem{Blaizot88}  J.P. Blaizot and G. Ripka, Phys. Rev. {\bf D38}, 1556
(1988).

\bibitem{Li87}  B.A. Li, K.F. Liu and M.M. Zhang, Phys. Rev. {\bf D35}, 1693
(1987).

\bibitem{Wambach87}  J. Wambach, H.W. Wyld, H.M. Sommermann, Phys. Lett. 
{\bf B186}, 272 (1987).

\bibitem{Hadjuk84}  Ch. Hajduk and B. Schwesinger, Phys. Lett. {\bf B140,}
172 (1984); {\bf B145}, 171 (1984); Nucl. Phys. {\bf A453}, 620 (1986).

\bibitem{Biedenharn85}  L.C. Biedenharn, Y. Dothan and M. Tarlini, Phys.
Rev. {\bf D31}, 649 (1985).

\bibitem{Braaten85}  E. Braaten and J.P. Ralston, Phys. Rev. {\bf D31}, 598
(1985).

\bibitem{Liu84}  K.F. Liu, J.S. Zhang and G.R.E. Black, Phys. Rev. {\bf D30}%
, 2015 (1984).

\bibitem{Schroers94}  B.J. Schroers, Z. Phys. {\bf C61,} 479 (1994)

\bibitem{Dorey94}  N. Dorey, J. Hughes and M.P. Mattis, Phys. Rev. {\bf D50}%
, 5816 (1994).

\bibitem{Adkins83}  G.S. Adkins, C.R. Nappi and E. Witten, Nucl. Phys. {\bf %
B228}, 552 (1983); G.S. Adkins, C.R. Nappi, Nucl. Phys. {\bf B233}, 109
(1984).

\bibitem{PDG}  Particle Data Goup, Phys. Rev. {\bf D54}, 1 (1996).

\bibitem{Sanyuk93}  V.G. Mankhankov, Y.P. Rybakov and V.I. Sanyuk, The
Skyrme model: fundamentals, methods, applications. Springer-Verlag, pages
120-123 (1993).
\end{references}
\end{document}